\documentstyle[12pt]{article}

\advance\hoffset by -7mm
\renewcommand{\title}[1]{\null\vspace{25mm}

\noindent{\Large{\bf #1}}\vspace{10mm}

\noindent {\large By }}
\newcommand{\authors}[1]{\noindent{\large #1}\vspace{20mm}

}
\newcommand{\address}[1]{\noindent #1\vspace{5mm}

}
\renewcommand{\abstract}[1]{\vspace{17mm}

\noindent{\small{\em Abstract.} #1}\vspace{2mm}

}
\newcommand{\journal}[4]{{\em #1~}{\bf #2}\,(19#3)\,#4.}

\newcommand{\hpa}{\journal {Helv. Phys. Acta}}

\newcommand{\ijmp}{\journal {Int. J. Mod. Phys.}}

\newcommand{\pr}{\journal {Phys. Rev.}}

\newcommand{\cmp}{\journal {Comm. Math. Phys.}}
\newcommand{\cqg}{\journal {Class. Quantum Grav.}}

\newcommand{\np}{\journal {Nucl. Phys.}}
\newcommand{\pl}{\journal {Phys. Lett.}}

\newcommand{\prep}{\journal {Phys. Reports}}

\makeatletter
\@addtoreset{equation}{section}
\makeatother

\newcommand{\lc}{\left[}\newcommand{\rc}{\right]}

\renewcommand{\a}{\alpha}
\renewcommand{\b}{\beta}
\renewcommand{\d}{\delta}
\newcommand{\e}{\varepsilon}
\newcommand{\f}{\phi}

\newcommand{\x}{\xi}
\renewcommand{\l}{\lambda} \renewcommand{\L}{\Lambda}
\newcommand{\m}{\mu}
\newcommand{\n}{\nu}

\newcommand{\r}{\rho}
 
\renewcommand{\t}{\theta}

\newcommand{\VV}{{\cal V}}

\newcommand{\XX}{{\cal X}}
\newcommand{\YY}{{\cal Y}}

\newcommand{\complex}{{\kern .1em {\raise .47ex
\hbox {$\scriptscriptstyle |$}}
    \kern -.4em {\rm C}}}
\newcommand{\real}{{{\rm I} \kern -.19em {\rm R}}}
\newcommand{\rational}{{\kern .1em {\raise .47ex
\hbox{$\scripscriptstyle |$}}
    \kern -.35em {\rm Q}}}

\newcommand{\dint}{\displaystyle{\int}}
\newcommand{\es}{\\[4mm]}

\newcommand{\pa}{\partial}

\newcommand{\fud}[2]{{\displaystyle{\frac{\delta #1}{\delta #2}}}}

\newcommand{\sla}{\raise.15ex\hbox{$/$}\kern -.57em}

\newcommand{\twiddle}{\lower.9ex\rlap{$\kern -.1em\scriptstyle\sim$}}

\newcommand{\equ}[1]{(\ref{#1})}

\newcommand{\eq}{\begin{equation}}
\newcommand{\eqn}[1]{\label{#1}\end{equation}}
\newcommand{\ee}{\end{equation}}
\newcommand{\eea}{\end{eqnarray}}
\newcommand{\eqa}{\begin{eqnarray}}
\newcommand{\eqan}[1]{\label{#1}\end{eqnarray}}
\newcommand{\ba}{\begin{array}}
\newcommand{\ea}{\end{array}}
\newcommand{\eqac}{\begin{equation}\begin{array}{rcl}}
\newcommand{\eqacn}[1]{\end{array}\label{#1}\end{equation}}

\def\6{\partial}
\def\={\!\!\!&=&\!\!\!}
\def\+{\!\!\!&&\!\!\!+~}

\begin{document}

\setcounter{page}{0}
\thispagestyle{empty}\hspace*{\fill} CERN-TH/95-29\\
\hspace*{\fill}TUW 95-04\\
\hspace*{\fill}hep-th 9502147\\
\title{The three-dimensional BF Model with Cosmological Term
in the Axial Gauge}
\authors{A. Brandhuber$^{\# 1}$, S. Emery$^{*2}$, K. Landsteiner$^{\#
1}$
  and M. Schweda$^{*}$}
\address{$^\#$ CERN, CH-1211 Gen{\`e}ve 23 (Switzerland)}
\address{$^*$ Institut f\"ur Theoretische Physik,
         Technische Universit\"at Wien\\
         Wiedner Hauptstra\ss e 8-10, A-1040 Wien (Austria)}
\abstract{\\
We quantize the three-dimensional $BF$-model using axial gauge
conditions. Exploiting the rich symmetry-structure of the model we
show that the Green-functions correspond to tree graphs and can be
obtained
as the unique solution of the Ward-Identities. Furthermore, we will
show
that the theory can be uniquely determined by symmetry considerations
without
the need of an action principle.}
\vspace{10mm}
\noindent CERN-TH/95-29\\
TUW 95-04\\
February 1995

\footnotetext[1]{Supported by the BMWF}
\footnotetext[2]{On leave from the D\'epartement de Physique
Th\'eorique,
  Universit\'e de Gen\`eve. Supported by the "Fonds Turrettini" and
  the "Fonds F. Wurth"}


\newpage
\section{Introduction}

Topological field models \cite{bbrt} of the Schwarz-type
\cite{schw-baku} have been the subject
of continous investigations over the recent years. These
theories are characterized by an invariant action which does not
depend on the metric structure of the manifold. Therefore, they are
devoid of any local observables. Nevertheless, the metric appears in
the
gauge-fixing term which is itself a BRST-variation.
The variation of the gauge-fixing term with respect to the metric is
a BRST-exact quantity implying the existence of a linear vector-like
supersymmetry \cite{dgs} $\n_\m$ in an elegant manner. Together with
the
BRST-symmetry, the symmetry $\n_\m$ forms an algebra of the
form
\eq
\{s,\n_\m\}=\pa_\mu
\eqn{topsusy}
stating that translations are no physical operations and thus
reflecting the topological nature of the theory. Thus one might say
that this relation lies at the heart of their topological properties.

The most prominent example of these theories is of course the
three-dimensional Chern-Simons theory which has led to the powerful
connection between link-invariants and the vacuum expectation value
of Wilson lines \cite{wit2}. The Chern-Simons theory has also been
studied extensively from a purely field theoretical point of view. It
turns out to be a completely ultraviolet finite theory and that this
finiteness is a direct consequence of the topological supersymmetry
\equ{topsusy}. Originally, this supersymmetry has been found using
the Landau-gauge \cite{dgs}. In a serie of papers it has been
generalized to other gauge-conditions as well \cite{blpss,gk,llss}.
Of particular interest was the case of the axial gauge where it
turned
out that
 the
topological supersymmetry is not only responsible for the finiteness
of
the
 theory,
but also allows to compute the Green functions without the use of
an action
principle \cite{blssep}. Let us also mention that it has been shown
that the topological supersymmetry exists also in string theories
\cite{ssw} and in two-dimensional chiral $W_3$-gravity \cite{cvcs}
and that it turned out to be an extremely useful tool for solving the
descent equations associated with the integrated BRST-cohomology
\cite{ws}.

Another class of Schwarz-type Topological theories~\cite{schw-baku}
are
the
BF models. Despite of their simple form they reveal a surprisingly
rich
symmetry
structure. Indeed, they allow for reducible invariances
\cite{luc-pig-sor}.
In the particular case of three dimensions, another interesting
feature
is
that the $B$-field is a $1$-form and therefore allows the addition of
a cubic term in $B$ into the action. This term is usually referred to
as a cosmological constant term since then the model is related to
three-dimensional Einstein-Hilbert gravity with such a term. A
detailed investigation of the symmetry stucture and finiteness
properties of this model in the Landau-gauge has been given in
\cite{ms}. Furthermore, the treatment of the same model on a manifold
with boundary in the axial gauge has been done in~\cite{nico}.

In the case of the Chern-Simons model in the axial gauge, we found
that
the
supersymmetry has two main consequences. Indeed, it turned out to be
strong
 enough
for fixing all the Green functions of the theory (in this sense, one
can say
 that it
can be substituted to the action principle) and it also imposes
the principal value prescription for the propagators.
Therefore, it  would be desirable to know wether this remains
valid for the BF system in the axial gauge. This is precisely the
question we
 will
address in this paper and we will show that the answer is
positive.

The work is organized as follows. In section 2 we introduce the
action and fix the notation and conventions. Section 3 presents the
symmetries and all the functional identities. We investigate their
consequences
 for
the equations of motion in section 4.1 and for the calculation of the
propagators in section 4.2. At the end, we propose some conclusion.

\section{The $3D$ BF model with cosmological term in the axial gauge}

The complete action of the $3D$ BF model containing a cosmological
term
with an axial gauge fixing is given by

\eq
S= S_{inv} + S_{gf}\,,
\eqn{action}
with
\eq
\ba{rl}
S_{inv}= & -\frac{1}{2} Tr \dint d^3x \left[ \epsilon^{\m\n\r}
 ( F_{\m\n} B_\r + \frac{2 \a}{3} B_\m B_\n B_\r ) \right]\,,  \es
S_{gf} =& Tr \dint d^3x \left[ b n^\m A_\m +
 d m^\m B_\m +
 \bar{c} n^\m (D_\m c + \a [ B_\m , \f]) +
    \bar{\f} m^\m (D_\m \f + [ B_\m , c])  \right]\,.
\ea
\eqn{split}

and $D_\m \ldots = \6_\m \ldots + g [ A_\m , \ldots ] $ denoting the
gauge covariant derivative. $F_{\mu\nu}$ is the  field strength of
the gauge field $A_\mu$. Further $b$, $d$ are the Langrange
multipliers imposing the gauge-conditions $n^\m A_\m = 0$ and $m^\m
B_\m= 0$ where $n^\m$ and $m^\m$ are {it a priori} two independent
gauge
fixing directions. $\bar{c}$, $c$ and $\bar{\f}$, $\f$ are the
anti-ghost and
ghost fields corresponding to the two gauge symmetries
of $S_{inv}$
\eq
\ba{rl}
\delta^{1} A_\mu =-D_\mu\theta\,, & \delta^{1} B_\mu =-\lc
B_\mu\,,\theta\rc\,, \es
\delta^{2} A_\mu =-\a\lc B_\mu,\l\rc\,,& \delta^{2} B_\mu =-D_\mu
\l\,.
\ea
\eqn{gaugesym}
We choose the gauge group to be simple, all
fields belong to the adjoint representation and are written as Lie
algebra matrices $\varphi(x) = \varphi^a(x) t_a$, with

\eq
    [t_a,t_b] = {f^c}_{ab} t_c ~, \qquad Tr(t_a t_b) = \d_{ab}.
\eqn{lie}
Finally $\a$ is some numerical constant.
We summarize the canonical dimensions and the ghost numbers of the
various fields in Table~\ref{dim}.

\begin{table}[hbt]
\centering
\begin{tabular}
{|l|            r|   r|   r|   r|   r|    c|   r|    c|    } \hline
         \vspace{-4mm} &&&&&&&& \\
             &$A$ &$B$ &$b$ &$d$ &$c$ &$\bar{c}$ &$\f$ &$\bar{\f}$ \\
\hline
Dimension    &$1$ &$1$ &$2$ &$2$ &$0$  &$2$  &$0$  &$2$    \\ \hline
Ghost number &$0$ &$0$ &$0$ &$0$ &$1$ &$-1$  &$1$  &$-1$   \\ \hline
\end{tabular}
\caption[t1]{Dimensions and ghost numbers.}
\label{dim}
\end{table}


\section{Symmetries of the action and Ward identities }

The action~\equ{action}
is invariant under the BRST transformation $s$ :
\eq\ba{ll}
sA_\m =- D_\m c - \a [B_\m,\f] ,\quad
& sB_\m =- D_\m \f - [B_\m,c] ,\es
s c = c^2 + \a \f^2 , & s \f = \{\f,c \}    , \es
s \bar{c} = b       , & s \bar{\f} = d      , \es
s b = 0             , & s d = 0             . \es
\ea\eqn{brs}
Since we are dealing with a topological field theory of Schwarz type,
the only metric dependence arises from the gauge fixing part of the
action. Therefore, the energy momentum tensor is BRST exact:
\eq
  T_{\a\b} = s \L_{\a\b}
\eqn{emtensor}
with
\eq\ba{ll}
 \L_{\a\b} = & Tr ( \eta_{\a\b} \bar{c} n^\r A_\r - \bar{c} n_\a A_\b
-
 \bar{c} n_\b A_\a + \eta_{\a\b} \bar{\f} n^\r B_\r - \bar{\f} n_\a
B_\b -
 \bar{\f} n_\b B_\a )  .
\ea\eqn{lambda}
Using the equations of motion, one gets for the divergence
of~\equ{lambda} the
following expression
\eq\ba{ll}
 \6^\a \L_{\a\b} = & Tr( \6_\b \bar{c} \fud{S}{b} - A_\b \fud{S}{c} -
 n^\a \bar{c} \e_{\r\b\a} \fud{S}{B_\r} +
 \6_\b \bar{\f} \fud{S}{b} - B_\b \fud{S}{\f} - \es
 & - m^\a \bar{\f} \e_{\r\b\a} \fud{S}{A_\r} +
 n^\a \bar{c} \e_{\r\b\a} m^\r d - n^\a \bar{\f} \e_{\r\b\a} m^\r b)
+ {\rm tot.\ der.}
\ea\eqn{div-emtensor}
Integrating~\equ{div-emtensor} allows to derive the usual form for
the
topological supersymmetry only for the case where\footnote{Actually
one
could
 also
insist in keeping different gauge vectors since the breaking term is
BRST exact. This breaking could be controlled by coupling it to an
additional
source and adding it to the action \cite{llss}} $n^\m=m^\m$ which we
will
assume for the rest of the paper.  Thus we have the following form
for
the
vector supersymmetry transformations $\n_\a$ :
\eq\ba{ll}
\n_\a A_\m =- \e_{\a\b\m} n^\b \bar{\f} ,\quad
& \n_\a B_\m =- \e_{\a\b\m} n^\b \bar{c} ,\es
\n_\a c =  A_\a     ,&
\n_\a \f =  B_\a    ,\es
\n_\a \bar{c} = 0       ,&
\n_\a \bar{\f} = 0      ,\es
\n_\a b = -\6_\a \bar{c} ,&
\n_\a d = -\6_\a \bar{\f} .
\ea\eqn{susy}
The  transformations $s$ and the supersymmetry transformations
$\n_\m$
form an algebra which closes on-shell:
\eq
s^2=\{\n_\m ,\n_\n\}=0,\quad \{s,\n_\m \}=\partial_\m
           + {\rm Eq.\ of\ motion}.
\eqn{onshell}\\
In addition there exist two discrete symmetries~\footnote{In
 fact,~\equ{discrete1}
and~\equ{discrete2} imply the existence of additional anti-BRST-like
symmetries
and anti-vector-like supersymmetries as in~\cite{dgs,blpss}.}
which leave the action \equ{action} invariant:
\eq\ba{lll}
 c \longleftrightarrow \a\bar{c} & , & \f \longleftrightarrow
\bar{\f}
\ea\eqn{discrete1}
and
\eq\ba{lll}
 c \longleftrightarrow \bar{\f} & , & \bar{c} \longleftrightarrow \f
\ea\eqn{discrete2}

At the level of the generating functional of the connected Green
functions
 $Z_C$,
all these symmetries leads to a set of WI. The one which correspond
to
the
 vector
supersymmetry takes the form
\eq\ba{ll}
\VV_\a Z_C = & Tr \dint d^3x \left( J_b \6_\a \fud{Z_C}{J_{\bar{c}}}
+
 \e_{\a\m\n} n^\m J_B^\n \fud{Z_C}{J_{\bar{c}}} + J_c
\fud{Z_C}{J_A^\a} +
 \right. \es & \left. +J_d \6_\a \fud{Z_C}{J_{\bar{\f}}} +
\e_{\a\m\n}
 n^\m J_A^\n \fud{Z_C}{J_{\bar{\f}}} + J_\f \fud{Z_C}{J_B^\a} \right)
= 0 .
\ea
\eqn{susyward}
In this formalism, the axial gauge is imposed by the two gauge
conditions :
\eq\ba{l}
 J_b + n^\m \fud{Z_C}{J_A^\m} = 0 ,\es
 J_d + n^\m \fud{Z_C}{J_B^\m} = 0 .
\ea\eqn{gaugecond}
As in any linear gauge there exist antighost equations which in the
case of  the axial gauge are local \cite{blpss}. In our case we have
two of
them and they take the following form:
\eq\ba{l}
 J_{\bar{c}} - n^\m \6_\m \fud{Z_C}{J_{c}} + \left[J_b,
\fud{Z_C}{J_{c}}
 \right] + \a \left[J_d, \fud{Z_C}{J_{\f}} \right] = 0 , \es
 J_{\bar{\f}} - n^\m \6_\m \fud{Z_C}{J_{\f}} + \left[ J_b ,
 \fud{Z_C}{J_{\f}} \right] + \left[J_d,\fud{Z_C}{J_{c}}
 \right] = 0 .
\ea\eqn{antighosteq}
Finally, it is well known that in the axial gauge, due to the
decoupling
of the ghosts, the Slavnov identity which express the invariance of
the
theory under the BRST-transformation~\equ{brs} takes the form of a
local WI.
Therefore, one get the two following local gauge WI's
\eq\ba{l}
 \6_\m J_A^\m - \left[J_A^\m, \fud{Z_C}{J_A^\m} \right] -
 \left[J_B^\m, \fud{Z_C}{J_B^\m} \right] -
\left[J_b,\fud{Z_C}{J_b}\right] -
 \left[J_d, \fud{Z_C}{J_d} \right] - \es
-\left\{J_c,\fud{Z_C}{J_c}\right\} -
 \left\{J_{\bar{c}},\fud{Z_C}{J_{\bar{c}}} \right\} -
 \left\{J_\f, \fud{Z_C}{J_\f} \right\} -
 \left\{J_{\bar{\f}}, \fud{Z_C}{J_{\bar{\f}}} \right\} +
 (n \cdot \6) \fud{Z_C}{J_b} = 0
\ea
\eqn{gaugeward1}
and
\eq\ba{l}
 \6_\m J_B^\m - \a \left[J_A^\m, \fud{Z_C}{J_B^\m} \right] -
 \left[J_B^\m, \fud{Z_C}{J_A^\m} \right] -
 \left[J_b, \fud{Z_C}{J_d} \right] -
 \a \left[J_d, \fud{Z_C}{J_b} \right] - \es -
 \a \left\{J_c, \fud{Z_C}{J_\f} \right\} -
 \left\{J_{\bar{c}}, \fud{Z_C}{J_{\bar{\f}}} \right\} -
 \left\{J_\f, \fud{Z_C}{J_c} \right\} -
 \a \left\{J_{\bar{\f}}, \fud{Z_C}{J_{\bar{c}}} \right\} +
 (n \cdot \6) \fud{Z_C}{J_d} = 0 .
\ea
\eqn{gaugeward2}


\section{Consequences of the Symmetries}

\subsection{Equations of motion}

Let us now investigate in some detail the meaning of the
relations of the last
section, starting with the projection of the WI for the
 supersymmetry\equ{susyward},
along the gauge fixed direction. Without loss
of generality we can choose the gauge vector $n^\m$ to be $(0,0,1)$.
We
will
denote this gauge fixed direction by $u$ and the transverse
coordinates
by $x^{tr} = (x^i), i=1,2$. Then,
\eq
\VV_u Z_C =  Tr \dint d^3x \left( J_b \6_u \fud{Z_C}{J_{\bar{c}}}
+ J_c  \fud{Z_C}{J_A^u} +J_d \6_u \fud{Z_C}{J_{\bar{\f}}} +
                       J_\f \fud{Z_C}{J_B^u} \right) = 0 .
\eqn{susyu}
Taking into accout the gauge conditions~\equ{gaugecond}, the latter
can
be
 written
as
$$
Tr \dint d^3x \left( J^b \XX + J^d\YY \right) = 0 .
$$
where $\XX$ and $\YY$ are the most general forms compatible
with~\equ{susyu}
$$
\ba{l}
\XX =  \6_u \fud{Z_C}{J_{\bar{c}}} +\l_1\left[ J^d,
 \fud{Z_C}{J_{\bar{c}}}\right]
+\l_2\left[ J^d, \fud{Z_C}{J_{\bar{\f}}}\right]
+\l_3\left[ J^b, \fud{Z_C}{J_{\bar{c}}}\right]
+\l_4\left[ J^b, \fud{Z_C}{J_{\bar{\f}}}\right]- J_c =0 \es
\YY =  \6_u \fud{Z_C}{J_{\bar{\f}}} +\x_1\left[ J^d,
 \fud{Z_C}{J_{\bar{c}}}\right]
+\x_2\left[ J^d, \fud{Z_C}{J_{\bar{\f}}}\right]
+\x_3\left[ J^b, \fud{Z_C}{J_{\bar{c}}}\right]
+\x_4\left[ J^b, \fud{Z_C}{J_{\bar{\f}}}\right]- J_\f =0
\ea
$$
At this level, one can use the consistency condition between the two
equations  we just found and the ghost equation~\equ{ghosteq} in
order
to fix the arbitrary parameters. Then, the result is
\eq\ba{l}
 J_c - n^\m \6_\m \fud{Z_C}{J_{\bar{c}}} + \left[J_b,
\fud{Z_C}{J_{\bar{c}}}
 \right] + \left[J_d, \fud{Z_C}{J_{\bar{\f}}} \right] = 0 , \es
 J_\f - n^\m \6_\m \fud{Z_C}{J_{\bar{\f}}} + \left[ J_b ,
 \fud{Z_C}{J_{\bar{\f}}} \right] + \a\left[J_d,\fud{Z_C}{J_{\bar{c}}}
 \right] = 0 .
\ea\eqn{ghosteq}
which are nothing else than the ghost equations. Thus, the
equations of motion for the ghost sector are a consequence of the
gauge-fixed component of the
supersymmetry WI, the gauge condition and the Slavnov identity.

For the gauge sector, let us consider the transverse component
of~\equ{susyward}
\eq\ba{r}
\VV_i Z_C =  Tr \dint d^3x \left( J_b \6_i \fud{Z_C}{J_{\bar{c}}}
+ \e_{ji} J_B^j \fud{Z_C}{J_{\bar{c}}} + J_c  \fud{Z_C}{J_A^i} +
\right. \es
\left. +J_d \6_i \fud{Z_C}{J_{\bar{\f}}} +
\e_{ji} J_A^j \fud{Z_C}{J_{\bar{\f}}} + J_\f \fud{Z_C}{J_B^i} \right)
= 0 .
\ea
\eqn{susytr}
together with the antighost equations~\equ{antighosteq} written as
functional
operators acting on $Z_C$
\eq\ba{l}
\left\{\6_u \fud{}{J_c} - \left[ J_b, \fud{}{J_c} \right]
     - \a \left[ J_d , \fud{}{J_{\f}} \right] \right\} Z_C =
J_{\bar{c}}, \es
\left\{\6_u \fud{}{J_{\f}} - \left[ J_b , \fud{}{J_{\f}} \right]
                 - \left[J_d,\fud{}{J_{c}} \right]\right\} Z_C =
J_{\bar{\f}} .
\ea\eqn{antighfun}
Their consistency gives rise to
the following identities
\eq\ba{l}
\left\{ \6_u \fud{}{J_A^i} - \left[ J_d, \fud{}{J_A^i} \right]
     - \a \left[ J_b, \fud{}{J_B^i} \right] \right\} Z_C
                 = \e_{ji} J_B^j-\6_i J_d, \es
\left\{ \6_u \fud{}{J_B^i} - \left[ J_d , \fud{}{J_B^i} \right]
                 - \left[ J_b,\fud{}{J_A^i} \right] \right\} Z_C
                  = \e_{ji} J_A^j-\6_i J_b .
\ea\eqn{gaugeeq}
which correspond to the equations of motion for the gauge fields.
This
concludes
the analysis of the consequences of the supersymmetry for the
equations
of
 motion.

\subsection{Calculation of the Propagators}

\subsection*{Gauge conditions}

We begin by looking at the gauge conditions \equ{gaugecond} which
imply
the vanishing of the connected Green functions involving the
components
$A_3$ or $B_3$
\eq\ba{l}
 \langle A_3^a(x) \prod_i \varphi_i(x_i) \rangle = 0 \es
 \langle B_3^a(x) \prod_i \varphi_i(x_i) \rangle = 0 \ea
                                   \qquad \forall \varphi_i
\eqn{B3X}
with two exceptions
\eq\ba{l}
 \langle A_3^a(x) b^b(y) \rangle = - \d^{ab} \d^{(3)}(x-y) , \es
 \langle B_3^a(x) d^b(y) \rangle = - \d^{ab} \d^{(3)}(x-y) .
\ea\eqn{B3d}

\subsection*{Antighost equations}

The antighost equations \equ{ghosteq} give the following differential
equations for the connected Green functions involving one pair of
ghost fields:
\eq
 \6_{x^3}\langle \bar{c}^a(x) c^b(y) \rangle = \d^{ab} \d^{(3)}(x-y)
,
\eqn{cbarceq}
\eq
 \6_{x^3}\langle \bar{\f}^a(x) \f^b(y) \rangle = \d^{ab}
\d^{(3)}(x-y)
\eqn{phibarphieq}
and
\eq\ba{l}
 \6_{x^3} \langle d^{c_1}(z_1) .. d^{c_r}(z_r) b^{b_1}(y_1) ..
 b^{b_r}(y_r) c^b(y)\bar{c}^a(x) \rangle = \es
 \displaystyle{\sum_{j=1}^s}
 f^{ab_jc}~\langle d^{c_1}(z_1) .. d^{c_r}(z_r)
 b^{b_1}(y_1) .. \widehat{b^{b_j}}(y_j) .. b^{b_r}(y_r)
 c^b(y)\bar{c}^c(y_j) \rangle~\d(x-y_j)~+ \es
 \displaystyle{\sum_{i=1}^r}
 f^{ac_ic}~\langle d^{c_1}(z_1)..\widehat{d^{c_i}}(z_i)..
 d^{c_r}(z_r) b^{b_1}(y_1)..b^{b_r}(y_r)
 c^b(y)\bar{\f}^c(z_i)\rangle~\d(x-z_i)
\ea\eqn{G1}
and
\eq\ba{l}
 \6_{x^3} \langle d^{c_1}(z_1) .. d^{c_r}(z_r) b^{b_1}(y_1) ..
 b^{b_r}(y_r) c^b(y)\bar{\f}^a(x) \rangle = \es
 \displaystyle{\sum_{j=1}^s}
 f^{ab_jc}~\langle d^{c_1}(z_1) .. d^{c_r}(z_r)
 b^{b_1}(y_1) .. \widehat{b^{b_j}}(y_j) .. b^{b_r}(y_r)
 c^b(y)\bar{\f}^c(y_j) \rangle~\d(x-y_j)~+ \es
 \a \displaystyle{\sum_{i=1}^r}
 f^{ac_ic}~\langle d^{c_1}(z_1)..\widehat{d^{c_i}}(z_i)..
 d^{c_r}(z_r) b^{b_1}(y_1)..b^{b_r}(y_r)
 c^b(y)\bar{c}^c(z_i)\rangle~\d(x-z_i) .
\ea\eqn{G2}
The solutions {\bf of} the equations \equ{cbarceq} and
\equ{phibarphieq}
are
\eq
 \langle \bar{c}^a(x) c^b(y) \rangle = \d^{ab} [ \t(x^3-y^3) + \b_1]
 \d^{(2)}(x^{tr} - y^{tr}) ,
\eqn{cbarc}
\eq
 \langle \bar{\f}^a(x) \f^b(y) \rangle = \d^{ab} [ \t(x^3-y^3) +
\b_2]
 \d^{(2)}(x^{tr} - y^{tr})
\eqn{phibarphi}\\
The form of the terms proportional to $\b_1$, $\b_2$ is dictated by
transverse two dimensional Poincar{\'e} invariance and scale
invariance.
Indeed, the latter forbids solutions of the type
$1/(x^{tr}-y^{tr})^2$
because this
term is not a well defined distribution. To give it a meaning would
need the introduction of UV subtraction point, i.e., of a
dimensionful
parameter which would break scale invariance.
The integration constant can be fixed with the help of the
discrete symmetry of the action \equ{discrete1} to be $\b_1 = \b_2 =
 -\frac{1}{2}$.\\
Integration of the equations \equ{G1} and \equ{G2} yields
the following recursion relations for the Green functions with
one pair of ghosts:

\eq\ba{l}
 \langle d^{c_1}(z_1) .. d^{c_r}(z_r) b^{b_1}(y_1) ..
 b^{b_s}(y_s) c^b(y)\bar{c}^a(x) \rangle = \es
 \qquad =\displaystyle{\sum_{j=1}^s}
 f^{ab_jc}~[\t(x^3-y_j^3) + \b(r,s)]\d^{(2)}(x^{tr} -
y_j^{tr})\times\\
 \qquad\times\langle d^{c_1}(z_1) .. d^{c_r}(z_r)
 b^{b_1}(y_1) .. \widehat{b^{b_j}}(y_j) .. b^{b_s}(y_s)
 c^b(y)\bar{c}^c(y_j) \rangle~+ \es
 \qquad +\displaystyle{\sum_{i=1}^r}
 f^{ac_ic}~[\t(x^3-z_i^3) + \b(r,s)]\d^{(2)}(x^{tr} -
z_i^{tr})\times\\
 \qquad\times\langle d^{c_1}(z_1)..\widehat{d^{c_i}}(z_i)..
 d^{c_r}(z_r) b^{b_1}(y_1)..b^{b_s}(y_s)
 c^b(y)\bar{\f}^c(z_i)\rangle
\ea\eqn{G1rec}
and
\eq\ba{l}
 \langle d^{c_1}(z_1) .. d^{c_r}(z_r) b^{b_1}(y_1) ..
 b^{b_s}(y_s) c^b(y)\bar{\f}^a(x) \rangle = \es
 \qquad =\displaystyle{\sum_{j=1}^s}
 f^{ab_jc}~[\t(x^3-y_j^3) + \b(r,s)]\d^{(2)}(x^{tr} -
y_j^{tr})\times\\
 \qquad\times\langle d^{c_1}(z_1) .. d^{c_r}(z_r)
 b^{b_1}(y_1) .. \widehat{b^{b_j}}(y_j) .. b^{b_s}(y_s)
 c^b(y)\bar{\f}^c(y_j)\rangle~+ \es
 \qquad +\a\displaystyle{\sum_{i=1}^r}
 f^{ac_ic}~[\t(x^3-z_i^3) + \b(r,s)]\d^{(2)}(x^{tr} -
z_i^{tr})\times\\
 \qquad\times\langle d^{c_1}(z_1)..\widehat{d^{c_i}}(z_i)..
 d^{c_r}(z_r) b^{b_1}(y_1)..b^{b_s}(y_s)
 c^b(y)\bar{c}^c(z_i)\rangle ~.
\ea\eqn{G2rec}\\
Using the discrete symmetry one could produce two additonal recursion
relations which are not written explicitly since they are not needed
for
our calculations.
The integration constants are all
fixed to $\b(r,s) = -\frac{1}{2}$ by the discrete symmetry
\equ{discrete1}
and Bose symmetry of the Lagrange multipliers $b$ and $d$.\\
Now we will discuss the recursion relations for some special values
of
$r$ and $s$.
\subsubsection*{The case $r=0$}
For this discussion we will use a more symbolic notion i.e. we
will drop indices and variables because we only want to find the
vanishing Green functions whereas the non vanishing Green functions
can always be obtained from the explicit recursion relations
\equ{G1rec} and \equ{G2rec}.
{}From \equ{G1rec} we get $\langle(b)^s c \bar{c} \rangle = \sum
\langle(b)^{s-1} c \bar{c} \rangle $ which ends up after $s$
steps with the $\langle c \bar{c} \rangle$ propagator defined in
\equ{cbarceq}. On the contrary $\langle(b)^s \f \bar{c} \rangle = 0$
since the recursion relation stops with the vanishing Green function
$\langle \f \bar{c} \rangle$. Using again the discrete symmetry we
obtain $\langle(b)^s \bar{\f} c \rangle = 0$ and $\langle(b)^s \f
\bar{\f} \rangle = \sum\langle(b)^{s-1} \f \bar{\f} \rangle $.
\subsubsection*{The case $s=0$}
In this case we have have to use \equ{G1rec} and \equ{G2rec}
iteratively e.g. $\langle(d)^r c \bar{c} \rangle = \sum
\langle(d)^{r-1} c \bar{\f} \rangle = \langle(d)^{r-2}c\bar{c}\rangle
=
\ldots $ . The final result
depends on wether the recursion relation ends up with $\langle c
\bar{c} \rangle$ which gives a non vanishing result or with $\langle
c \bar{\f}
\rangle$ which gives zero. Here we only want to compile the
zero results:\\
$\langle(d)^r c \bar{c} \rangle =0$ and $\langle(d)^r
\f \bar{\f} \rangle =0$ if $r$ is an odd integer.\\
$\langle(d)^r \f \bar{c} \rangle =0$ and $\langle(d)^r
c \bar{\f} \rangle =0$ if $r$ is an even integer.\\

The Green functions with additional ghosts, gauge fields or Lagrange
multipliers vanish in general as a consequence of the antighost
equations \equ{ghosteq}:
\eq
 \langle X c\bar{c} \rangle =
 \langle X c\bar{\f} \rangle =
 \langle X \bar{c}\f \rangle =
 \langle X \bar{c} \bar{\f}\rangle = 0 ~, ~{\rm unless}~
 X = (b)^m (d)^n ~.\eqn{ghpair}

\subsection*{Transverse supersymmetry}

{}From equation \equ{susytr} we get further relations along the
same
lines. For reasons of simplification we use the sloppy
notation from the discussion above whenever possible.\\
The results for the two-point functions are:
\eq
 \langle A^a_i(x) B^b_j(y) \rangle = \e_{ij}\d^{ab} [\t(x^3 - y^3)
 -\frac{1}{2}]\d^{(2)}(x^{tr}-y^{tr}) ~,\eqn{AB}
\eq
 \langle b^a(x) A^b_i(y) \rangle = -\d^{ab} [\t(x^3 - y^3)
 -\frac{1}{2}]\6_{x^i}\d^{(2)}(x^{tr}-y^{tr}) ~,\eqn{bA}
\eq
 \langle d^a(x) B^b_i(y) \rangle = -\d^{ab} [\t(x^3 - y^3)
 -\frac{1}{2}]\6_{x^i}\d^{(2)}(x^{tr}-y^{tr}) \eqn{dB}
and $\langle AA \rangle = \langle BB \rangle = \langle bb \rangle =
\langle dd \rangle =\langle dA \rangle =\langle bB \rangle =
\langle c\bar{\f} \rangle = \langle \bar{c}\f \rangle = 0$.
Furthermore we observe that all two-point functions with one ghost
and
one bosonic field vanish.\\
For the higher Green functions we obtain the recursion relations
\eq\ba{c}
\langle b^{d_1}(w_1)..b^{d_r}(w_r)d^{c_1}(z_1)..d^{c_s}(z_s)
A^{b_1}_{l_1}(y_1)..A^{b_t}_{l_t}(y_t)
B^{a_1}_{n_1}(x_1)..B^{a_u}_{n_u}(x_u)
\left\{ \ba{c} A^b_i(y)\\B^b_i(y) \ea\right\} \rangle = \es
=\displaystyle{\sum_{k=1}^r}
\6_{w_k^i} \langle b^{d_1}(w_1)..\widehat{b^{d_k}}(w_k)..b^{d_r}(w_r)
(d)^s(A)^t(B)^u \left\{ \ba{c} c^b(y)\\ \f^b(y) \ea\right\}
\bar{c}^{d_k}(w_k)\rangle~+ \es
+\displaystyle{\sum_{k=1}^s}\6_{z_k^i}
\langle (b)^rd^{c_1}(z_1)..\widehat{d^{c_k}}(z_k)..d^{c_s}(z_s)
(A)^t(B)^u \left\{ \ba{c} c^b(y)\\ \f^b(y) \ea\right\}
\bar{\f}^{c_k}(z_k)\rangle~+ \es
+\displaystyle{\sum_{k=1}^t} \e_{il_k}\langle (b)^r(d)^s
A^{b_1}_{l_1}(y_1)..\widehat{A^{b_k}_{l_k}}(y_k)..A^{b_t}_{l_t}(y_t)
(B)^u \left\{ \ba{c} c^b(y)\\ \f^b(y) \ea\right\}
\bar{\f}^{b_k}(y_k)\rangle~+ \es
+\displaystyle{\sum_{k=1}^u} \e_{in_k}\langle (b)^r(d)^s(A)^t
B^{a_1}_{n_1}(x_1)..\widehat{B^{a_k}_{n_k}}(x_k)..B^{a_u}_{n_u}(x_u)
\left\{ \ba{c} c^b(y)\\ \f^b(y) \ea\right\} \bar{c}^{b_k}(y_k)\rangle
.\es
\ea\eqn{susyrec}
In the following we want to specify these recursion relations for
special values of $r$, $s$, $t$ and $u$ to demonstrate that all Green
functions can be obtained from our recursion relations. All Green
functions with one pair of ghosts have been obtained in the previous
subsection. Now we want to calculate the remaining Green functions
with only bosonic fields and at least one $A$ or $B$ field.

\subsubsection*{The case $t=u=0$}
We obtain from \equ{susyrec}:
$$\langle(b)^r(d)^sA\rangle = \sum
\6\langle(b)^{r-1}(d)^sc\bar{c}\rangle+\sum\6\langle(b)^r(d)^{s-1}c
\bar{\f}\rangle
$$
so the calculations breaks down to summing over
already known Green functions. The same holds for
$\langle(b)^r(d)^sB\rangle $ .

\subsubsection*{The case $r=s=0$}
We obtain
$$\langle(A)^t(B)^uA\rangle =
\e\langle(A)^t(B)^{u-1}c\bar{c}
\rangle + \e\langle(A)^{t-1}(B)^uc\bar{\f}\rangle
$$
Using~\equ{ghpair}
we find $\langle (A)^a(B)^b \rangle = 0$ unless $a=1$ and $b=1$ which
yields the two-point function $\langle AB \rangle$ .

\subsubsection*{The case $s=u=0$}
In this case the relation \equ{susyrec} takes the form:
$$\langle
(b)^r(A)^t \left( \ba{c} A \\ B \ea\right) \rangle = \sum\6
\langle (b)^{r-1}(A)^t \left( \ba{c}c \\ \f \ea\right) \bar{c}\rangle
+
\sum\e\langle(b)^r(A)^{t-1}\left( \ba{c} c\\ \f
\ea\right)\bar{\f}\rangle
$$
For $t=0$ we get the results:\\
$\langle (b)^r A \rangle = \sum\6 \langle (b)^{r-1}c \bar{c}\rangle$
and $\langle (b)^r B \rangle = 0$\\
For $t=1$:\\
$\langle (b)^rAB \rangle = \sum\e\langle(b)^r \f \bar{\f}\rangle$ and
$\langle (b)^rAA \rangle = 0$.

\subsubsection*{The case $r=u=0$}
In this case the relation \equ{susyrec} takes the form:
$$
\langle
(d)^s(A)^t \left( \ba{c} A \\ B \ea\right) \rangle = \sum\6
\langle (d)^{s-1}(A)^t \left( \ba{c}c \\ \f \ea\right)
\bar{\f}\rangle +
\sum\e\langle(d)^s(A)^{t-1}\left( \ba{c} c\\ \f
\ea\right)\bar{\f}\rangle
$$
For $t=0$ we get the results:\\
$\langle (d)^s A \rangle = 0$ for $s$ odd
and $\langle (d)^s B \rangle = 0$ for $s$ even.\\
For $t=1$:\\
$\langle (d)^sAA \rangle = 0$ for $s$ even and
$\langle (d)^sAB \rangle = 0$ for $s$ odd.

\subsubsection*{The case $r=t=0$}
In this case the relation \equ{susyrec} takes the form:
$$
\langle
(d)^s(B)^u \left( \ba{c} A \\ B \ea\right) \rangle = \sum\6
\langle (d)^{s-1}(B)^u \left( \ba{c}c \\ \f \ea\right)
\bar{\f}\rangle +
\sum\e\langle(d)^s(B)^{u-1}\left( \ba{c} c\\ \f
\ea\right)\bar{c}\rangle
$$
For $t=0$ we get the results:\\
$\langle (d)^s A \rangle = 0$ for $s$ odd
and $\langle (d)^s B \rangle = 0$ for $s$ even.\\
For $t=1$:\\
$\langle (d)^sBA \rangle = 0$ for $s$ odd and
$\langle (d)^sBB \rangle = 0$ for $s$ even.

\subsection*{Gauge invariance}

The two gauge symmetries \equ{gaugeward1} and \equ{gaugeward2} do not
give a lot of new information besides consistency checks and the fact
that all correlators consisting only of the Lagrange multipliers
vanish:
$$
 \langle b^{a_1}(x_1)..b^{a_m}(x_m)d^{b1}(y_1)..d^{b_n}(y_n)
 \rangle = 0 \qquad \forall~m,~n
$$
This is the unique solution obeying dimensional and scaling
arguments.


\section{Concluding Remarks}
The main result of our study is that the Green functions of the model
are the unique solutions of the Ward-identities defining
the theory. Furthermore it turned out that the topological vector
supersymmetry imposed a rather unexpected restriction on the
{\it a priori} independent gauge vectors.
It is also worth noticing that the Green functions correspond to tree
graphs only. Note also that in principle there are loop graphs with
external $b$ and $d$ fields only, however, as in Chern-Simons theory
\cite{mar}, the gauge-field and the ghost field contributions to
these graphs cancel exactly due to the topological supersymmetry.
Having investigated here the three-dimensional BF model it is now
natural to apply the axial gauge also to higher dimensional
BF models.
It would be highly interesting to know wether the methods developped
here
and in \cite{blssep} are also applicable to these cases.


\section{Acknowledgements}

The authors would like to thank O. Piguet for helpful
discussions and comments.



\end{document}